\documentclass[a4paper,11pt]{article}
\usepackage{jinstpub} 
\usepackage{lineno}
\usepackage{hyperref}
\usepackage{mathtools}

\title{Modeling Scintillation Photon Transport and Reconstruction Algorithms for the Time-of-Flight Detector in the T2K Neutrino Experiment}

\author[h]{C.~Alt}
\author[a]{A.~Blanchet}
\author[a]{S.~Bordoni}
\author[a,j]{P.~Collard}
\author[b]{T.~H.~Bui}
\author[c]{M.~H.~Bui}
\author[b]{G.~Ha}
\author[g]{C.~Jesús-Valls}
\author[a]{V.~S.~Kasturi}
\author[k]{A.~Klustová}
\author[a]{A.~Korzenev\thanks{now at Joint Institute for Nuclear Research, Dubna, Russia.}}
\author[e]{T.~A.~Le}
\author[f]{T.~Lux}
\author[b]{A.~D.~Nguyen}
\author[b]{D.~T.~Nguyen}
\author[b]{H.~Nguyen}
\author[a]{S.~Samani}
\author[a]{F.~Sanchez}
\author[b]{M.~Ta}
\author[i]{T.~Thaiduc}
\author[g,a]{E.~Villa}

\affiliation[a]{Département de Physique Nucléaire et Corpusculaire, Université de Genève, Geneva, Switzerland}

\affiliation[b]{Faculty of Physics, VNU University of Science,
Hanoi, Vietnam}

\affiliation[c]{IOP, Vietnam Academy of Science and Technology,
Hanoi, Vietnam}

\affiliation[e]{INST, Vietnam Atomic Energy Institute,
Hanoi, Vietnam}

\affiliation[f]{Institut de Física d’Altes Energies (IFAE), The Barcelona Institute of Science and Technology (BIST), Campus UAB, E-08193 Bellaterra, Barcelona, Spain}

\affiliation[g]{CERN, European Organisation for Nuclear Research, CH‑1211 Genève 23, Switzerland.}

\affiliation[h]{Institute for Particle Physics and Astrophysics, ETH Zürich, CH‑8093 Zürich, Switzerland}

\affiliation[i]{Boston University, Boston, USA}

\affiliation[j]{Universit\'{e} Claude Bernard Lyon 1, Facult\'{e} des Sciences, D\'{e}partement de Physique, Villeurbanne, France}

\affiliation[k]{Imperial College London, Department of Physics, Blackett Laboratory, SW7 2BW London, United Kingdom}

\emailAdd{federico.sancheznieto@unige.ch}

\abstract{
The T2K ND280 upgrade aims to reduce the systematic uncertainty of the CP-violating phase, $\delta_{CP}$, to reject non-CP violation hypothesis at $3\sigma$ confidence level. A crucial component of the ND280 upgrade, alongside the Super Fine Grained Detector (SuperFGD) and two High-Angle Time Projection Chambers (TPCs), is the Time-of-Flight (ToF) detector, which significantly enhances background rejection and particle identification capabilities. The ToF detector features six modules in a cube configuration, each with 20 plastic scintillator bars measuring $\text{220}\times\text{12}\times\text{1}\,\text{cm}^3$ and is equipped with Silicon Photomultiplier (MPPC) arrays at both ends to capture scintillation light. This letter outlines the modelling of the detector response and the signal reconstruction process. }

\keywords{ Time-of-Flight, SiPM, MPPC, scintillators, T2K, neutrino physics }

\begin{document}
\maketitle
\flushbottom
\section{Introduction}
\label{sec:intro}
In recent years, the precision demands of neutrino oscillation experiments have stretched the capabilities of existing near detectors. These detectors are crucial for calibrating beam neutrino predictions and reducing uncertainties in neutrino-nucleus cross-sections. To address limitations in angular coverage and granularity for low-momentum particles, T2K initiated an upgrade in 2016. This upgrade also aimed to enhance particle direction determination to minimise background noise from outside the detector's active fiducial volume. The improvements are designed to optimize systematic uncertainties, handle increasing data volumes, and enhance the precision of $\delta_{CP}$ measurements, enabling rejection of the non-CP-violation hypothesis at approximately the $3\sigma$ confidence level \cite{T2K:2019bbb,T2K:2019bcf}.

A significant part of this upgrade involved replacing the P0D detector with a system featuring the Super Fine Grained Detector (SuperFGD)~\cite{Blondel:2020hml}, positioned between two High Angle Time Projection Chambers (HA-TPCs)~\cite{Attie:2022smn} enclosed by six Time-of-Flight (TOF) detectors. The SuperFGD acts as a neutrino target, while the TOF detectors measure particle arrival times to distinguish between external particles and those from neutrino interactions, thereby reducing background signals.

This paper details the reconstruction algorithm and the modelling of photon transport and electronics response, focusing on a simple yet physically descriptive model. This approach allows for the use of tabulated values and an understanding of each simulation element's impact. The development of both algorithms was carried on and tested using data from the single bar setup \cite{Korzenev:2021mny}, a test bench consisting of a single detector unit.


In this paper, we first introduce the study (Section~1) and present the ToF detector (Section~2). We then describe the methods for waveform analysis (Section~3) and particle position and time reconstruction (Section~4). Section~5 details the detector modelling, including photon processes and electronics, while Section~6 compares the model with experimental data. Finally, Section~7 summarizes the main conclusions.

\section{ToF detector description}
\label{sec:Detector}
\begin{figure}
    \centering
    \includegraphics[width=0.5\linewidth]{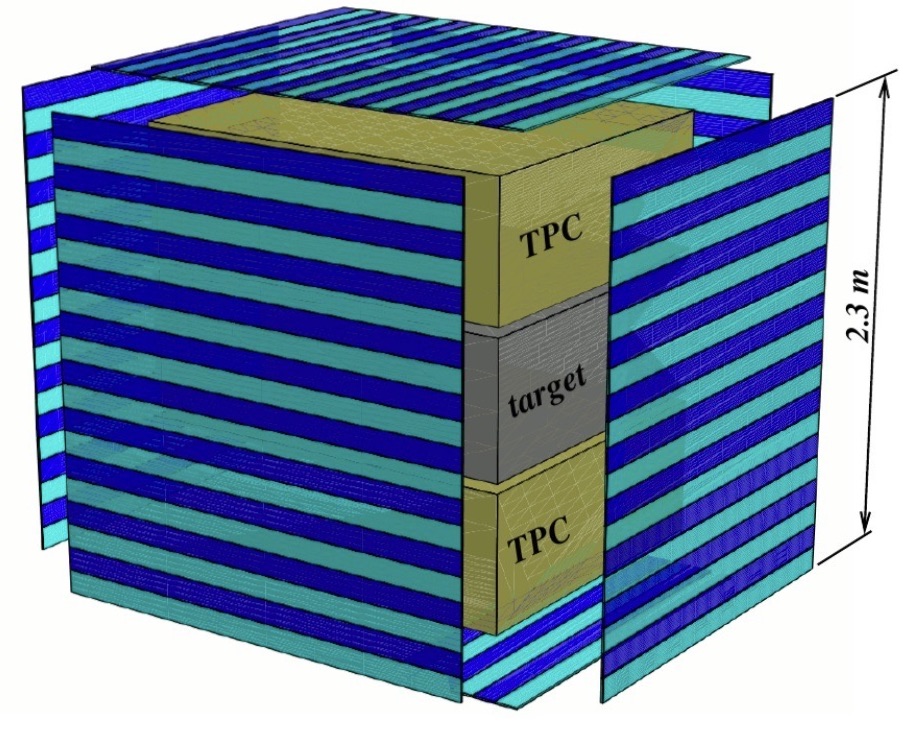}
    \caption{3D view of the ToF system, composed of six modules arranged in a cube. Each module houses 20 plastic scintillator bars, covering an active area of 5.4 m². The bars are mounted in a plane with 1.5 mm gaps to accommodate steel brackets on the aluminium frame. In the bottom module, two bars are removed to provide space for cable trays.}
    \label{fig:3DToF}
\end{figure}

The ToF system is made up of six similar modules arranged in a cube, see Fig.~\ref{fig:3DToF}. Each module contains 20 plastic scintillator bars, covering an active area of 5.4 m². Due to space constraints and the need for a sturdy build, the bars are placed in a plane with a 1.5 mm gap between them, allowing for steel brackets to attach the bars to an aluminium frame. Two bars from the bottom module are removed to make room for cable trays.

Each bar measures $\text{220}\times\text{12}\times\text{1} \text{cm}^3$ and is made of EJ-200 plastic scintillator, which has a fast rise ($0.9$~ns) and decay time ($2.1$~ns) and a peak emission wavelength ($420$~nm) that matches the spectral sensitivity of the MPPCs. The bars are wrapped in aluminium foil and black plastic film to ensure opacity. The aluminium maximises specular reflection along the bar, enhancing light collection efficiency and reducing photon arrival time dispersion, a feature used in the simulation.

\begin{figure}
    \centering
\includegraphics[width=0.6\linewidth]{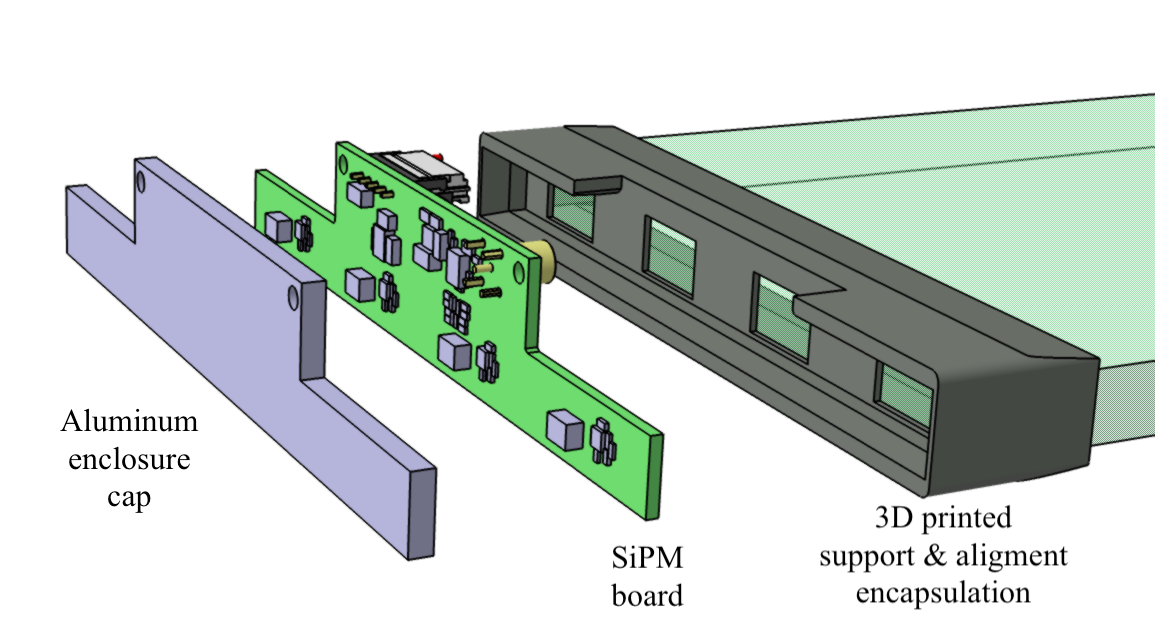}
    \includegraphics[width=0.6\linewidth]{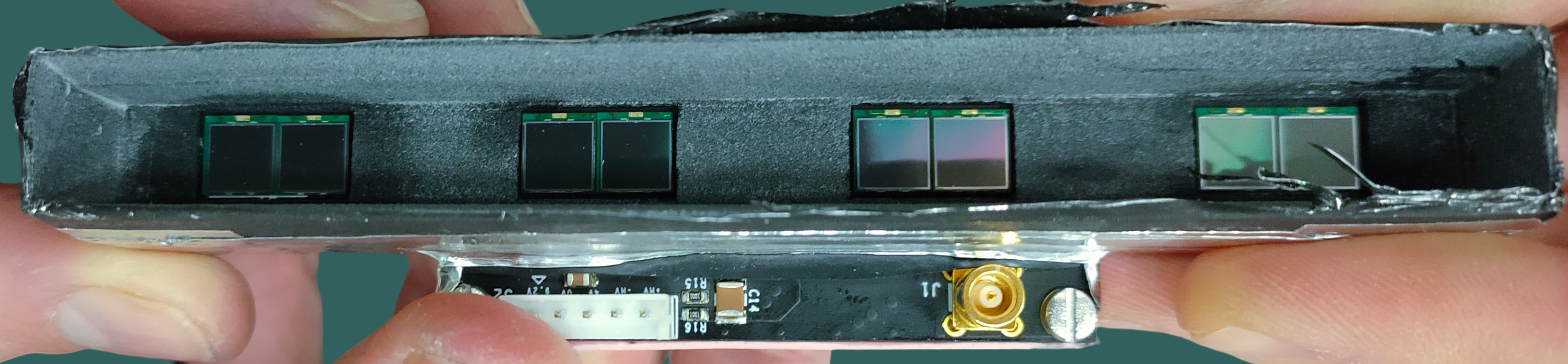}
    \caption{Layout of MPPC sensors and 3D printed protective capsule at the readout end of the bar. }
    \label{fig:PictureEndBar}
\end{figure}

Each scintillator bar is read out at both ends by an array of 8 Hamamatsu S13360-6050PE MPPCs (6 × 6 mm² each), with the signals from each array summed into a single readout channel, yielding 2 channels per bar and a total of 236 channels across the detector (two bars from the bottom layers are removed to allow cabling access). The MPPC arrays cover approximately 24\% of the active area at each end of the bar, which is important when considering reflections: photons missing the photosensors can reflect off the bar ends and contribute to the detected signal. These MPPC arrays are housed in plastic end-caps to ensure mechanical stability and light-tightness, see Fig.~\ref{fig:PictureEndBar}.

\subsection{Single bar setup}

\begin{figure}
    \centering
    \includegraphics[width=1.0\linewidth]{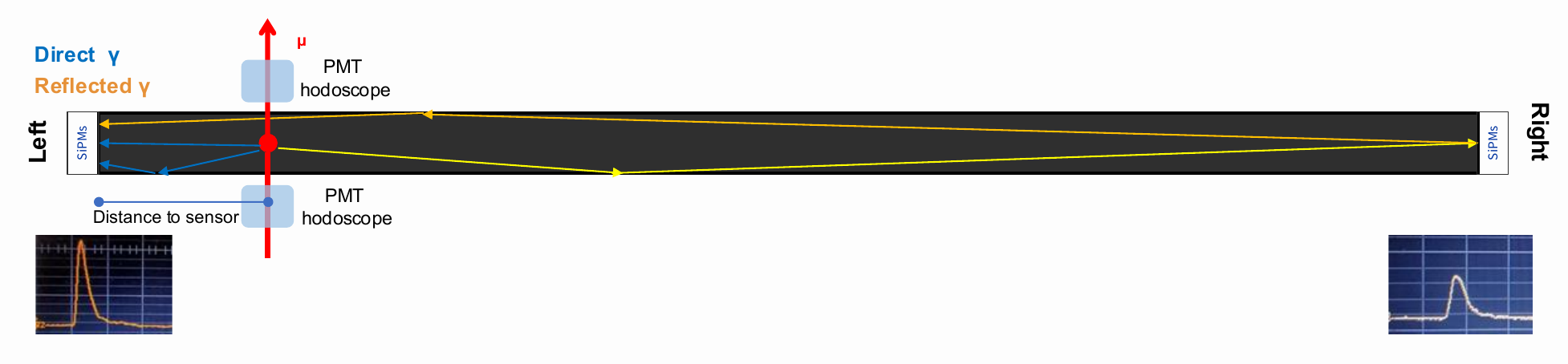}
    \caption{Sketch of the single-bar test-bench. The trajectories of photons reaching the sensor directly (blue and yellow) and after reflection at the far end (orange) are shown, together with the two MPPC readout ends (Left and Right) and the two reference photomultiplier hodoscopes. The expected waveforms from both readout ends are also illustrated.  }
    \label{fig:SgBarSketch}
\end{figure}

The performance of the system was evaluated using a single ToF bar, see sketch in Fig.~\ref{fig:SgBarSketch}. The experimental setup mirrored the final detector geometry and readout configuration including the final SAMPIC digitiser~\cite{SAMPIC} that is used in the T2K 4$\pi$ Time of Flight detector~\cite{Korzenev:2021mny}\footnote{In this earlier work the same detector and front-end electronics was used with different waveform digitiser.}. In the single bar setup, two identical counters were positioned at the top and bottom of the bar to create a trigger system. Both counters were constructed from fast scintillator material (EJ-228) and were connected to two $1$-inch R4998 Hamamatsu photomultiplier tubes (PMTs) via light guides. The trigger time was determined by averaging the times recorded by the four PMTs. Signals from both the PMTs and the MPPC arrays of the test bar were directed to a SAMPIC digitiser for data acquisition. This electronic readout provided comprehensive waveform data over a 19.75~ns time interval with 62 samples of 0.3125~ns each\footnote{The SAMPIC readout provides 64 samples, although the final two are not sufficiently stable.}. Measurements were conducted at 14 distinct positions along the bar, achieved by repositioning the time reference hodoscopes.

\section{Detector modelling}
\label{sec:Modeling}
For its usage in the T2K's N280 detector, we have developed a detector model capable of describing the photon transport and electronics performance in detail using a simple algorithm based on fundamental physical inputs. The simulation accounts for photon propagation, photon detection, and the electronics response. 
In this section, we describe the components of the detector modelling, the model steering parameters are presented in Table~\ref{tab:SimParam}.

\begin{table}[h!]
\centering
\begin{tabular}{ll|ll}
\textbf{Parameter} & \textbf{Value} & \textbf{Parameter} & \textbf{Value} \\ \hline
\multicolumn{4}{c}{Geometry and optics parameters}\\
\hline
Bar length (cm) & 220. & Scintillator index of refraction & 1.58 \\ 
Silicon index of refraction~\cite{ReflInxDB} & 5.13 & Silicon cover index of refraction~\cite{Hamamatsu}  & 1.55 \\ 
Reflectivity MPPC house material & 0.3 & $g_{\text{Mie}}$ & 0.9 \\ 
$\Lambda_{\text{Mie}}$~(cm) & 10000. & $\Lambda_{\text{Rayl}}$~(cm) & 360. \\ 
$\lambda_{\text{Abs}}$~(cm) & 340.  & Scintillator raise time (ns)~\cite{SCIONIX} & 0.9 \\ 
Scintillator falling Time (ns)~\cite{SCIONIX} & 2.1 & Quantum efficiency~\cite{Hamamatsu} & 0.5 \\ 
Fill factor & 0.74 & Photo detector coverage & 0.24 \\ 
Silicon coverage & 0.313 & Photons per MeV~\cite{SCIONIX} & 10000 \\ 
Amplitude normalization & 0.0091 & --- & --- \\
\hline
\multicolumn{4}{c}{Readout electronics simulation} \\
\hline
Sampling time (ns)~\cite{SAMPIC} & 0.3125 & Noise level (V) & 0.0017 \\ 
Threshold (V) & 0.01 & Delay between MPPC (ns) & 0.21 \\ 
Total readout samples~\cite{SAMPIC} & 62 & Number of pre-samples & 15 \\
\hline
\multicolumn{4}{c}{Single Photon Wave Form parameters} \\
\hline
$\text{C}_0$ & 0.01 & $\text{C}_1$ & 20.0 \\ 
$\text{C}_2$ & 0.27 & $\text{C}_3$ & 2.1 \\ 
$\text{C}_4$ & 0.24 & $\text{C}_5$ & 2.7 \\ 
\hline
\end{tabular}
\caption{ Simulation parameters for the single-bar setup. The definitions and origins of these values are provided in the main text.}
\label{tab:SimParam}
\end{table}

\subsection{Photon generation} 
After a particle deposits energy in the detector, the number of generated photons is given by the following formula:
\begin{equation}
    N_{\gamma} = \frac{\Delta E}{F_{\gamma}}
\end{equation}
where $F_{\gamma}$ is specified by the scintillator manufacturer and equals $10,000 \, \gamma/\text{MeV}$~\cite{SCIONIX}.

Photons direction are generated uniformly within a spherical volume and subsequently propagated according to the relevant physical processes. At the time of generation, the scintillator response time is explicitly taken into account. This time is sampled on a photon-by-photon basis and modelled by a distribution given by the convolution of exponential components. In practice, it is parametrised as
\begin{equation}
g(t) = \left(1 - e^{-\frac{t}{t_{\text{rise}}}}\right) e^{-\frac{t}{t_{\text{fall}}}},
\label{Eq:DeExcitation}
\end{equation}
where $t_{\text{rise}} = 0.9$~ns denotes the characteristic rise time, and $t_{\text{fall}} = 2.1$~ns the fall time of the scintillation process~\cite{SCIONIX}.

\subsection{Photon transport: scattering and attenuation}
Since the bar is wrapped with a reflective layer, we assume perfect specular reflection for photon propagation. This simplifies the simulation, as reflections only increase the effective photon path length in the scintillator:
\begin{equation}
    d_{\gamma} = \frac{d_S}{\cos\theta}
\end{equation}
where $ d_S $ is the straight-line distance from the energy deposition point to the sensor. This assumption is justified because the reflective wrapping preserves the photon incidence angle upon reflection on the bar surfaces facing the readout sensors, so that reflections do not modify the photon’s angle of incidence when it reaches the readout surface. As a result, reflections do not alter the probability for photons to exit the bar at the ends and be detected by the photo-sensors. This is even more justified since, in the simulation, we do not consider the exact exit point of the photon at the readout end.

For photons reaching the sensor, we assign an arrival time:
\begin{equation}
t_{\gamma} = t_0 + t_{\text{scint}} + \frac{d_{\gamma}}{\eta \, c}
\end{equation}
where $ c $ is the speed of light, $ \eta $ is the refractive index, $ t_0 $ is the time of energy deposition and $t_{\text{scint}}$ is the scintillator response time distributed as given in Eq.~\ref{Eq:DeExcitation}.

During photon transport, three distinct scattering processes are considered, each characterised by a specific attenuation length and angular dispersion of the scattered photons.  Scintillation photons are assumed to have wavelength distribution centred around the nominal SiPM absorption frequency, wavelength-dependent variations in these parameters are neglected, and a single, wavelength-independent response is adopted for all photons.
\begin{itemize}

\item {\bf Rayleigh Scattering} is characterised by an interaction length $\lambda_{\text{Rayl}}$, which defines the mean free path of photons before undergoing a scattering event. 
Rayleigh scattering occurs off particles much smaller than the wavelength of light ($\lambda$), typically $< \lambda/10$.
The angular distribution of the scattered photon relative to its incident propagation direction is described by the scattering angle $\theta_{\text{Rayl}}$. The probability density for $\theta_{\text{Rayl}}$ follows the Rayleigh differential cross section:
\begin{equation}
p(\theta_{\text{Rayl}}) \propto 1 + \cos^{2}\theta_{\text{Rayl}} 
\end{equation}
The value of $\lambda_{Rayl}$ is not provided in the product specifications. We adopted a value on the order of the detector light attenuation length, as reported in~\cite{SCIONIX}.  It is noteworthy that the angular distribution of Rayleigh scattering is symmetric for forward and backward directions. As a result, the effect on photon arrival at the sensor is a temporal broadening: early photons emitted toward the sensor may be lost, while late photons originally emitted away from the sensor are redirected and detected.  In this sense, effective Rayleigh scattering can partially accommodate deviations from pure specular reflection in the photon’s propagation along the bar.

\item {\bf Mie Scattering} is instead described by an interaction length $\lambda_{\text{Mie}}$, which determines the average distance between successive Mie interactions. Mie scattering occurs for particles with sizes in the range $\lambda/10 < \text{particle size} < \lambda$, with $\lambda$ the photon wavelength. Unlike Rayleigh scattering, the angular distribution of the scattered photons is strongly forward-peaked and cannot be expressed in a simple analytic form. It is commonly parametrised using approximations such as the Henyey–Greenstein function~\cite{Ehlers04052023}:
\begin{equation}
p(\cos \theta_{\text{Mie}}) \propto \frac{1 - g^{2}}{\left(1 + g^{2} - 2g\cos\theta_{\text{Mie}}\right)^{3/2}}
\end{equation}
where $\theta_{\text{Mie}}$ is the scattering angle and $g$ is the asymmetry parameter, with $g \rightarrow 1$~\cite{refg} corresponding to strongly forward-directed scattering.

\item {\bf Absorption} is governed by an interaction length 
$\lambda_{Abs}$, representing the mean free path for photon absorption. Once absorbed, a photon is removed from the transport process and does not undergo further scattering or re-emission. Absorption can be split into passive and active processes. Passive absorption occurs as described above, while active absorption involves the excitation of scintillator molecules by incident photons. These excited molecules rapidly undergo non-radiative de-excitation, followed by the emission of lower-energy photons in accordance with the Stokes shift principle. The re-emitted photons have longer wavelengths and minimal overlap with the SiPM absorption spectrum. In practice, the effects of active and passive absorption are similar, and since our simulation does not track photon wavelength, we treat both processes in a unified way.

\end{itemize}
In practice, Mie scattering has a negligible impact on photon transport due to its strongly forward-peaked angular distribution and the typically large values of $\lambda_{\text{Mie}}$~\cite{refg}. Consequently, it is effectively disabled in fast simulation by assigning an extremely large value to $\lambda_{\text{Mie}}$, making the probability of a Mie interaction vanishingly small. In contrast, Rayleigh scattering contributes significantly to the randomisation of photon arrival times due to its symmetric angular distribution. This effect becomes particularly evident for photons travelling long distances, where it noticeably broadens the detected waveform shape.

Photon transport is simulated by propagating photons in steps sampled from an exponential distribution:
\begin{equation}
P(s) \propto e^{-\frac{s}{\lambda}}
\end{equation}
where the effective interaction length $\lambda$ is given, ignoring the Mie scattering, by
\begin{equation}
\frac{1}{\lambda} = \frac{1}{\lambda_{\text{Rayl}}} + \frac{1}{\lambda_{\text{Abs}}}
\end{equation}

If the photon exits the scintillator bar during this step, reflection and detection are handled according to the algorithms described below. If the photon remains within the bar, the type of interaction is selected probabilistically, with the probabilities for Rayleigh scattering, 
and absorption given by
\begin{equation}
P_{\text{Rayl}} = \frac{\lambda}{\lambda_{\text{Rayl}}}, \quad
P_{\text{Abs}} = \frac{\lambda}{\lambda_{\text{Abs}}}
\end{equation}
and the corresponding interaction is executed accordingly.

This should be interpreted as an effective parametrization. Imperfections in the specular reflection within the scintillator bar can be modelled either as an increased absorption probability or as an enhanced Rayleigh scattering rate. The numerical values employed in the simulation are tuned to reproduce the observed results and do not necessarily correspond to the exact physical parameters of the system.

\subsection{Photon reflection}

We also account for the reflection of photons at the far end of the bar. Photons emitted in the direction opposite to the sensor can be reflected with a given probability. This effect is simulated using an accurate description of the detector geometry in this region, see Fig.~\ref{fig:PictureEndBar}. 

The coupling between the bar and the SiPM is through an air gap of a few micrometers, typically between 1 and 10~$\mu$m.
To account for reflections at the interface, we use the Fresnel reflectance formula for unpolarised light:
\begin{equation}
    R =
    \left(
        \frac{n_i \cos\theta_i - n_f \cos\theta_f}{n_i \cos\theta_i + n_f \cos\theta_f}
    \right)^2
    +
    \left(
        \frac{n_i \cos\theta_f - n_f \cos\theta_i}{n_i \cos\theta_f + n_f \cos\theta_i}
    \right)^2
    \label{eq:Fresnel}
\end{equation}
which includes the contributions from both polarisation components of the light. 

In this formula, $n_i$ ($n_f$) is the refractive index of the initial (final) medium, and $\theta_i$ ($\theta_f$) is the angle of incidence in the initial (final) medium. The angle $\theta_f$ is determined from Snell’s law:
\begin{equation}
    n_i \sin\theta_i = n_f \sin\theta_f
\end{equation}
We consider two scenarios for the reflected photons:
\begin{itemize}
    \item When the photon hits the active region of the SiPM, we include three reflections: plastic--air, air--SiPM cover, and SiPM cover--SiPM. The photon angle is not altered since we assume perfect specular reflection in the silicon. The indices of refraction used are shown in Table~\ref{tab:SimParam} (air is taken as nominal 1). 
    \item When the photon falls outside the SiPM active area, we include reflections at the plastic--air and air--holder interfaces, as the 3D-printed PA12+CF30 (polyamide + $30$\% Carbon Fibre) support for the SiPMs is also exposed to scintillation light. In this case, the photon angle is regenerated isotropically to model the diffusive character of the rough surface. The reflectivity was approximately set to 0.3 to achieve good data-simulation agreement, with this value guided by experimental results~\cite{Material3DPrinted}.

\end{itemize}

The refractive index of the scintillator is provided by the manufacturer~\cite{SCIONIX}, while that of the cover material is given by its producer~\cite{Hamamatsu}. The refractive index of silicon at the typical absorption wavelength of the SiPM ($\approx 450$~nm) is taken from tabulated data~\cite{ReflInxDB}. 


\subsection{Photon detection}
Photon detection is simulated by computing the probability that a photon is absorbed in the SiPM. This probability is determined by three main factors:
\begin{itemize}
\item \textbf{Sensor coverage:} The detection probability scales with the ratio of the total sensor area to the cross-sectional area of the bar. This is represented by a factor $f_{Fill}$.
    
\item \textbf{Internal reflection:} Due to the change of the index of refraction from the scintillator to the air and then to the SiPM, a fraction of the photons will be reflected following the probabilities computed in Eq.~\ref{eq:Fresnel}.
    
\item \textbf{Quantum Efficiency:} The sensor has an intrinsic quantum efficiency for photons entering the SiPM. The manufacturer does not provide this value directly, but instead quotes a total photon detection efficiency (including surface reflectivity), which is specified to be below 40\% at 420~nm~\cite{Hamamatsu}. This value includes the inefficiencies due to the SiPM pixel fill factor, defined as the ratio of the active photo-sensor area to the total area. In our model, this total efficiency is taken as the product of the photon transmission through the surface (approximately 69\% for perpendicularly incident photons) and the intrinsic quantum efficiency, which we fix to 50\% (see Table~\ref{tab:SimParam}). This results in a combined efficiency of 34.5\%. Although slightly lower than the tabulated 40\%, this choice provides a reasonable description of the data, likely compensating for some deficiencies in the simulation, particularly in the tabulated photon yield per MeV provided by the scintillator manufacturer~\cite{SCIONIX}.
\end{itemize}

The internal reflection probability should not be simplified in the calculations because the arrival time depends on the angle~$\theta$. As a result, it has a larger effect on late-arriving photons than on early ones, which may impact the time resolution performance.  In practice, we adjust these parameters while ensuring that the number and timing of photons arriving at the sensor remain consistent with the results obtained from the single bar setup.

\begin{figure}
    \centering
    \includegraphics[width=0.49\linewidth]{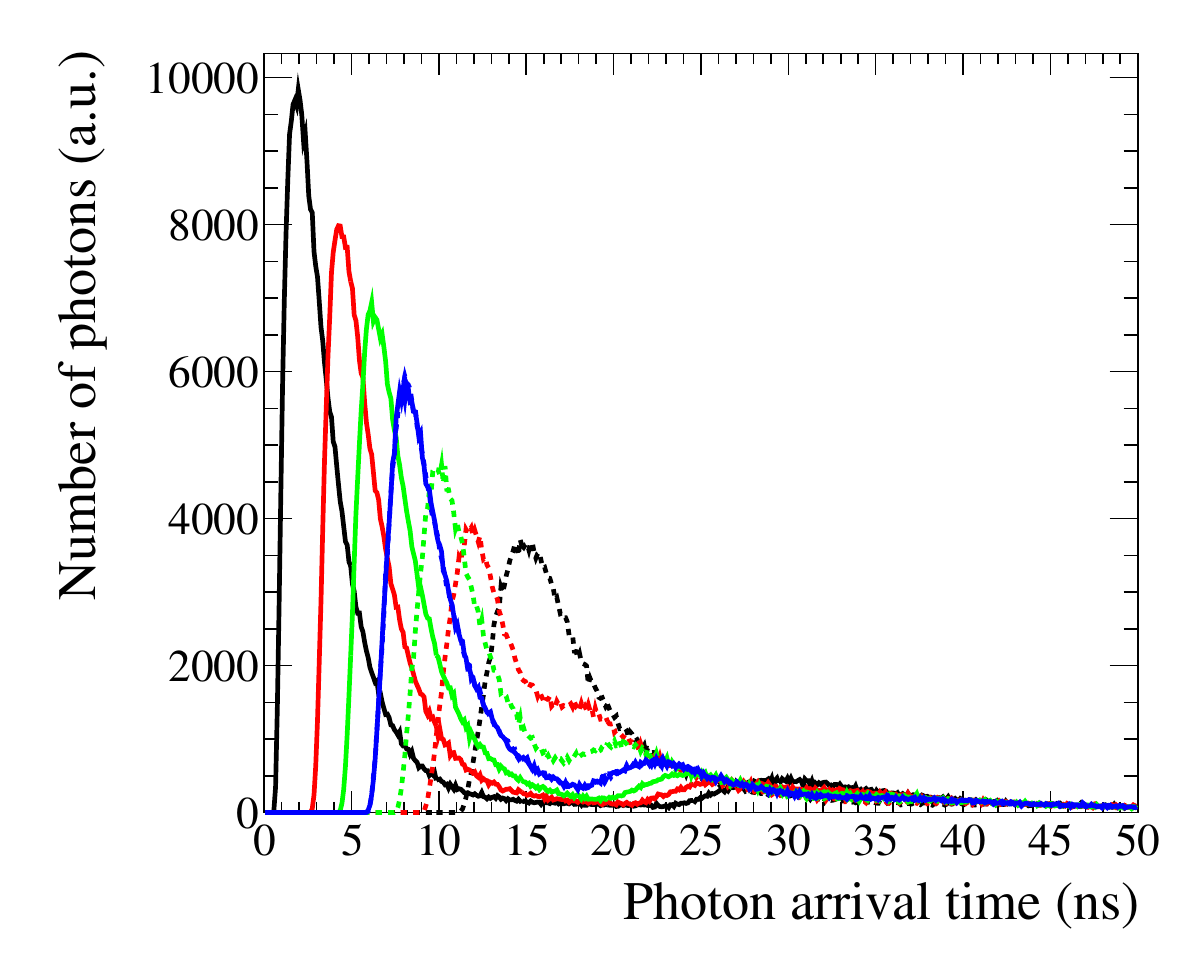}
    \includegraphics[width=0.49\linewidth]{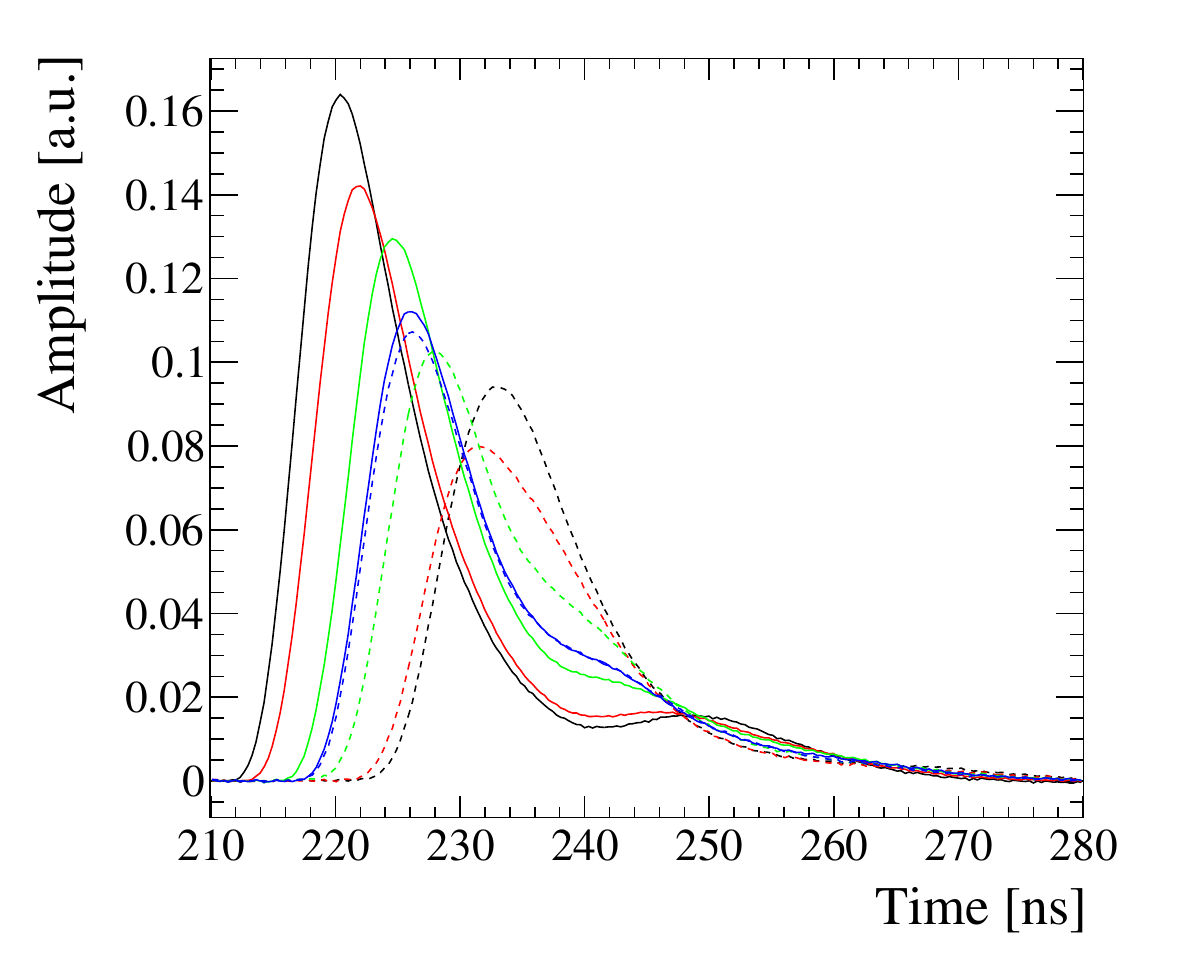}
    \caption{Left: Photon arrival times to the closer sensor (continuous line) and the far one (dashed line). The arrival times are shown for several minimal distances: 10~cm (black), 50~cm (red), 80~cm (green) and the middle of the bar (blue). Right: Waveforms were obtained using an alternative WaveCatcher system~\cite{7097545} capable of recording up to 1024 samples. The results are not directly comparable to the MC results since they include electronics signal processing.}
    \label{fig:PhotonArrivalTime}
\end{figure}

\begin{figure}
\centering
\includegraphics[width=0.75\linewidth]{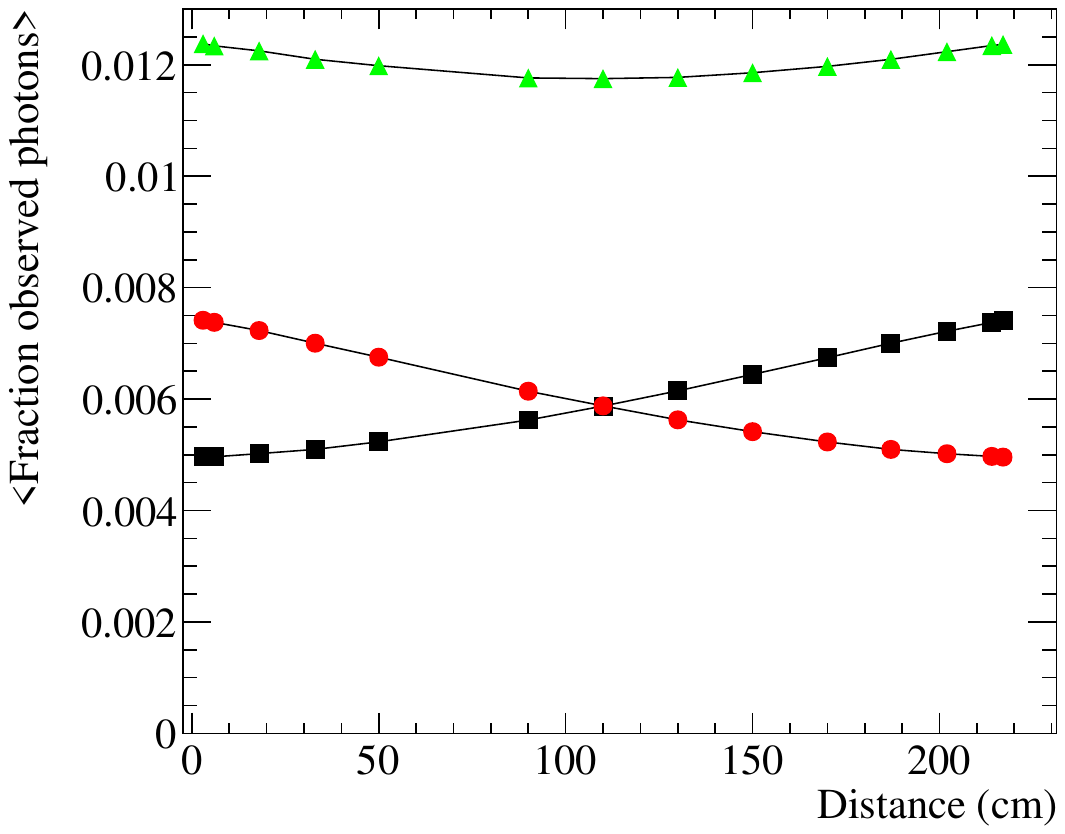}
\caption{Fraction of detected photons in the detector for each end as predicted by the model. Red dots shows photons detected at the sensor closer to the deposition, black squares for the opposite end and green triangles show the sum of both sensor ends.}
\label{fig:FracPhotons}
\end{figure}

Before proceeding with the implementation of the readout electronics, the qualitative behaviour of the detector can be inferred from the detected photon statistics. The photon time-of-flight (ToF) distributions for different source-to-sensor distances are illustrated in Fig.~\ref{fig:PhotonArrivalTime}. The observed temporal dispersion is on the order of few ns for the first arrival peak increasing with the distance, with two distinct peaks corresponding to the direct and reflected light components. As predicted, the temporal separation between the two peaks decreases monotonically with increasing distance to the readout sensor. At short distances, the peaks are well-resolved, while the peaks converge and eventually overlap near the distal end of the measurement range.

Additionally, the simulation enables the estimation of both the mean photon count and the relative photon flux observed at each detection endpoint as a function of the source-sensor separation, see Fig.~\ref{fig:FracPhotons}. The simulation predicts that the total number of photons read by both ends is 264 for a 2.2~MeV deposited energy. The sum of photons collected by the two end readouts reaches a maximum (0.0124\%) when the particle passes near the bar edges and decreases by approximately 6\% when it crosses at the centre. 

\subsection{Electronics response and digitization}
The electronics response consists of two components: the combined SiPM and front-end response, and the digitization process. The SiPM and front-end electronics response are measured in the laboratory using a picosecond laser and a Wavecatcher 12-bit 3.2-GS/s fast digitiser~\cite{7097545}, under the assumption that all photons arrive simultaneously. We parametrize the measured response with a function that reproduces the observed shape:
\begin{eqnarray}
f_{FE}(t) = \begin{cases} \text{C}_0 \, e^{ 
  - \frac{(t - \text{C}_1)^2}{ 
    2 \left[ \text{C}_2 (t - \text{C}_1) + \text{C}_3 \right]^2 }}, \forall\; t \le \text{C}_1 \\
    \text{C}_0 \, e^{ 
  - \frac{(t - \text{C}_1)^2}{ 
    2 \left[ \text{C}_4 (t - \text{C}_1) + \text{C}_5 \right]^2 }}, \forall\; t > \text{C}_1 
    \end{cases}
\end{eqnarray} 
This function can be interpreted as a Gaussian centred at $\text{C}_1$, the peak time of the distribution, with a width that varies linearly with time: $\text{C}_2 (t - \text{C}_1) + \text{C}_2$. We allow for an asymmetric response around the maximum by splitting the signal below or above $C_1$ with a different width dependency parametrised with $\text{C}_4$ and $\text{C}_5$. The values obtained from the laser measurements are shown in Table~\ref{tab:SimParam}.

The readout board is composed of eight sensors arranged in blocks of two, each separated by approximately 3~cm. Since the signals are summed into a single output, there is a potential difference in arrival times among them. This is accounted for by a constant time addition of $ n \cdot t_{\text{ROdelay}} $, where $ n \in \{0, 1, 2, 3\} $ and $ t_{\text{ROdelay}} \approx 210$~ps\footnote{We assume the main difference is due to the distance between sensors and a signal propagation speed in the PCB traces of 14~cm/ns.}, each with a probability of 25\%.

The measured response function is normalized to unity and convolved with the individual photon arrival times $t_{\gamma}$ to obtain the resulting waveform shape $W(t)$:
\begin{eqnarray}
W(t) = \sum_{\forall \gamma} f_{FE}(t - t_{\gamma})
\end{eqnarray}
Digitization is modelled as an analogue pipeline, where the charge is integrated over each sampling interval. 
The summed photon signal in each sample is then scaled by a calibration constant $N_{V}$, adjusted to match the values obtained with the single bar, to produce the output voltage.
Additionally, electronic noise is added, modelled as a normal distribution with width $\sigma_{Noise}$, also derived from the single bar experimental results.

After digitization, the time at which the signal crosses a predefined threshold is identified, and $ N_{\text{presamples}} $ samples preceding this point are recorded, along with a total of $ N_{\text{samples}} $ samples. Since the particle arrival times are not synchronized with the digitization clock, a random smearing within one clock period is applied to account for this effect. In our SAMPIC readout electronics~\cite{SAMPIC} we select $\tau_{\text{Sampling}} = 0.3125 $~ns.

\section{Waveform Parametrization and Time Reconstruction }
\label{Sec:WaveFormParam}
The waveform output provided by the SAMPIC electronics~\cite{SAMPIC} allows for a precise determination of time beyond the sampling time of the electronics. To take full advantage of the waveform resolution, each waveform is fitted with a modified Gaussian function: 
\begin{eqnarray}
 f(t) = \begin{cases}
  A_0 e^{-\frac{(t-A_1)^2}{2 (|A_2|(t-A_1)+A_3)^2 }}+A_6  & \forall\; t \leq A_1 \\
  A_0 e^{-\frac{(t-A_1)^2}{2 (|A_4|(t-A_1)+A_5)^2 }}+A_6  &  \forall\; t \geq A_1 \\
\end{cases}
\label{eq:WaveFormFit}
\end{eqnarray}
$A_0$ denotes the signal amplitude, $A_1$ the position of the maximum among all signal samples, and $A_6$ the channel electronics baseline. The Gaussian function is modified by introducing a linearly dependent standard deviation, $|A_{2,4}|(t - |A_1|) + A_{3,5}$, which differs for the rising and falling edges of the waveform. The parameters $A_3$ and $A_5$ characterise the Gaussian dependence of the rising and falling edges, respectively, while $A_2$ and $A_4$ parametrise the non-Gaussian deviations in the rising and falling edge of the waveform.

\begin{figure}
    \centering
    \includegraphics[width=0.49\linewidth]{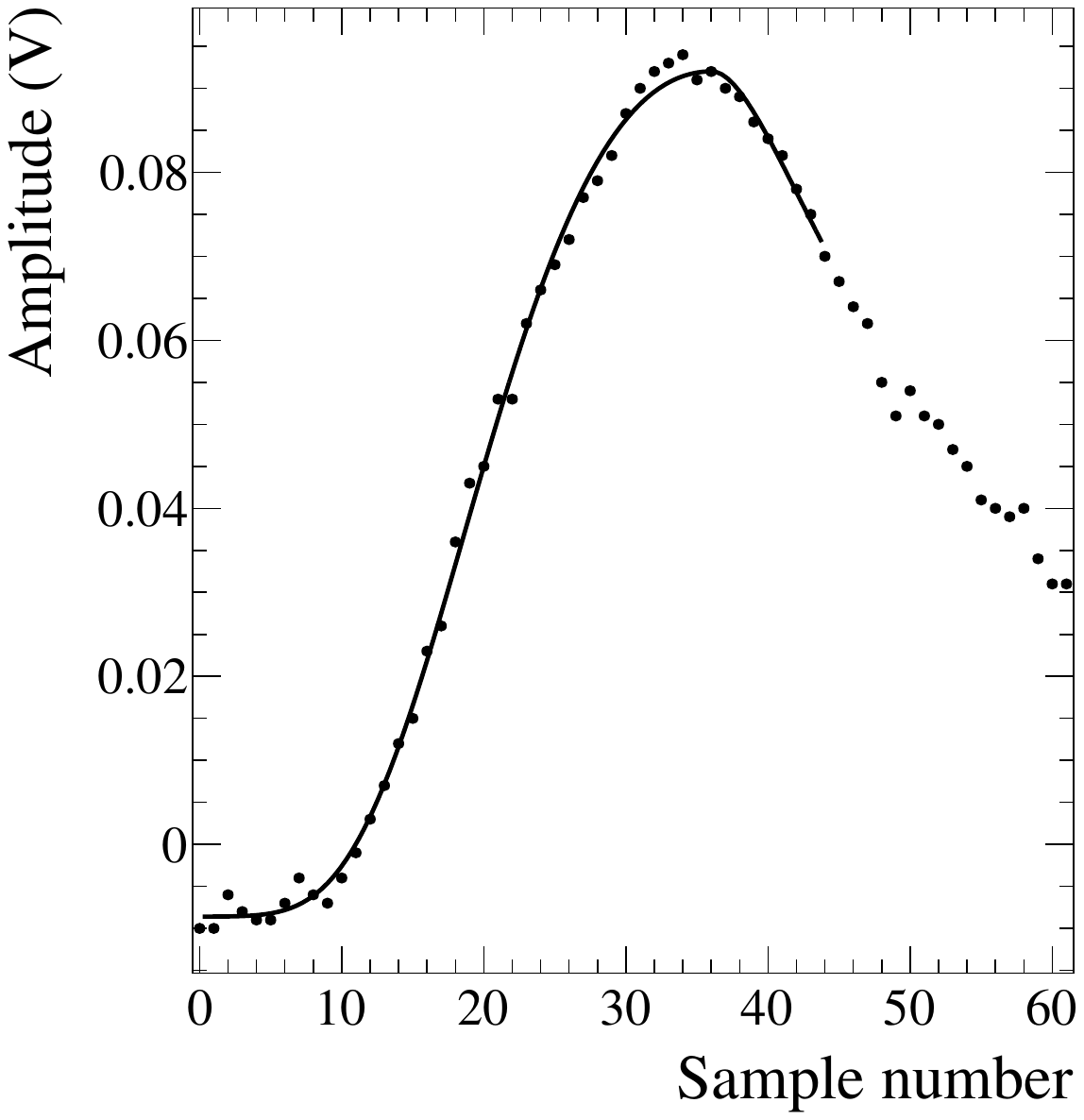}        \includegraphics[width=0.49\linewidth]{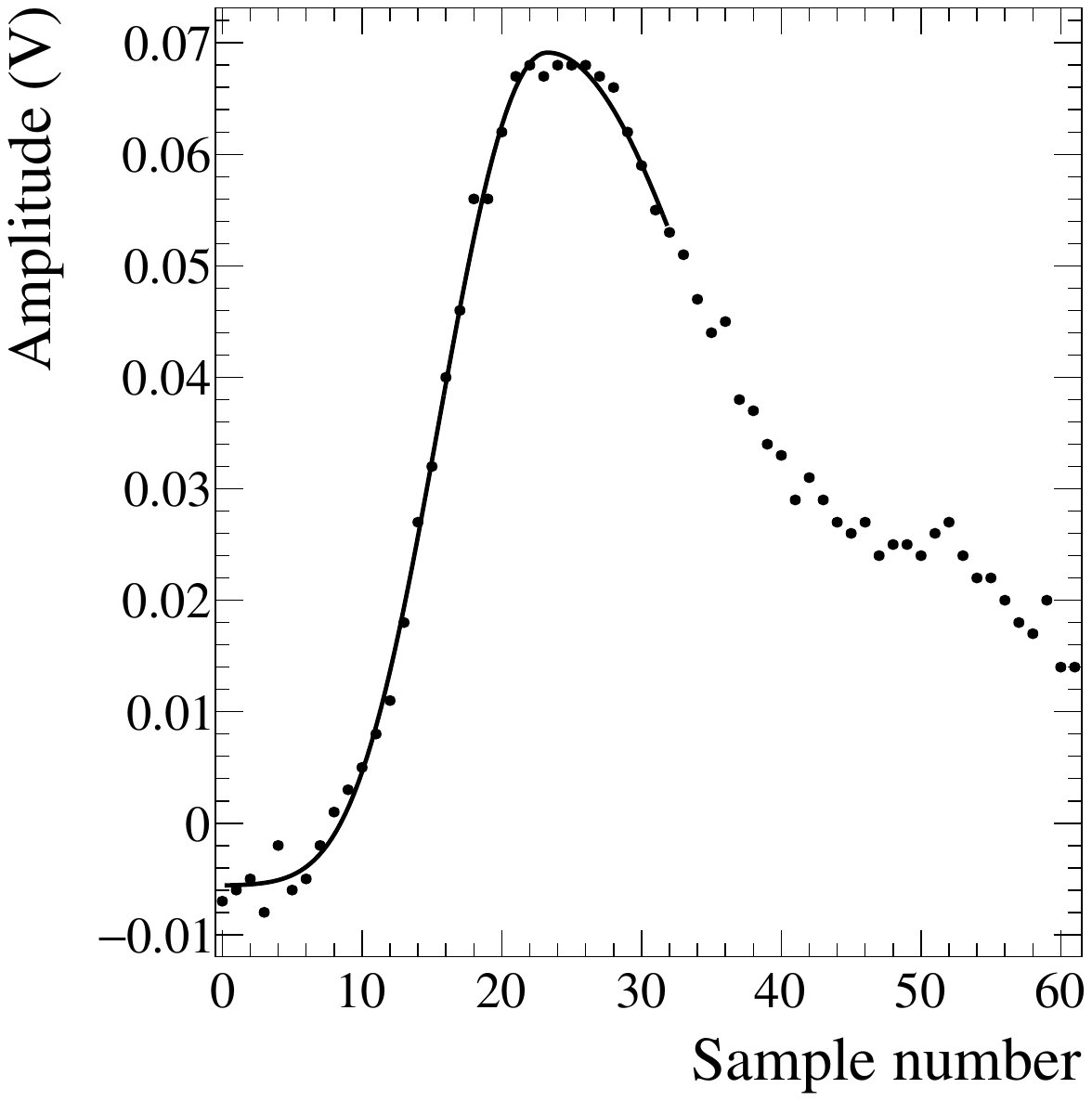}
    \caption{Two examples of real data waveforms fitted with the proposed waveform model are shown. Each bin represents the nominal readout sampling period of 0.3125~ns. }
    \label{fig:WFfit}
\end{figure}

The fit is restricted to all the samples before the maximum and 10 samples after the peak. The main reason is that the falling edge of the waveform does not provide relevant information for the timing, except that it helps to determine the waveform peak position. The falling edge also depends strongly on the superposition of the photons reflected at the other end of the bar, making the fit strongly dependent on the bar position. A more detailed discussion of the $A_i$ parameter dependencies with the distance is provided in Sec.~\ref{Sec:DataSimComp}, following the presentation of the photon propagation simulation in Sec.~\ref{sec:Modeling}. The quality of the fit adjustment is shown for two cases in Fig.~\ref{fig:WFfit}.

The waveform time is determined by setting a fraction $\tau$ of the maximum amplitude $A_0$ and solving the equation:
\begin{eqnarray}
  f(t) - A_6 = \tau\, A_0,\,  \forall\, t \le A_1 
\end{eqnarray}
leading to the solution of a second order polynomial : 
\begin{eqnarray}
\left(A_2^2 + \frac{1}{2 \ln{\tau}}\right)(t-A_1)^2 + 2 |A_2|\,A_3\,(t-A_1) + A_3^2 = 0    
\end{eqnarray}

This method, which condenses the waveform information into a compact formula, has proven to be stable and reliable, yielding slightly better results than standard constant-threshold or local-interpolation methods applied to the raw waveform. 

\begin{figure}
    \centering
    \includegraphics[width=\linewidth]{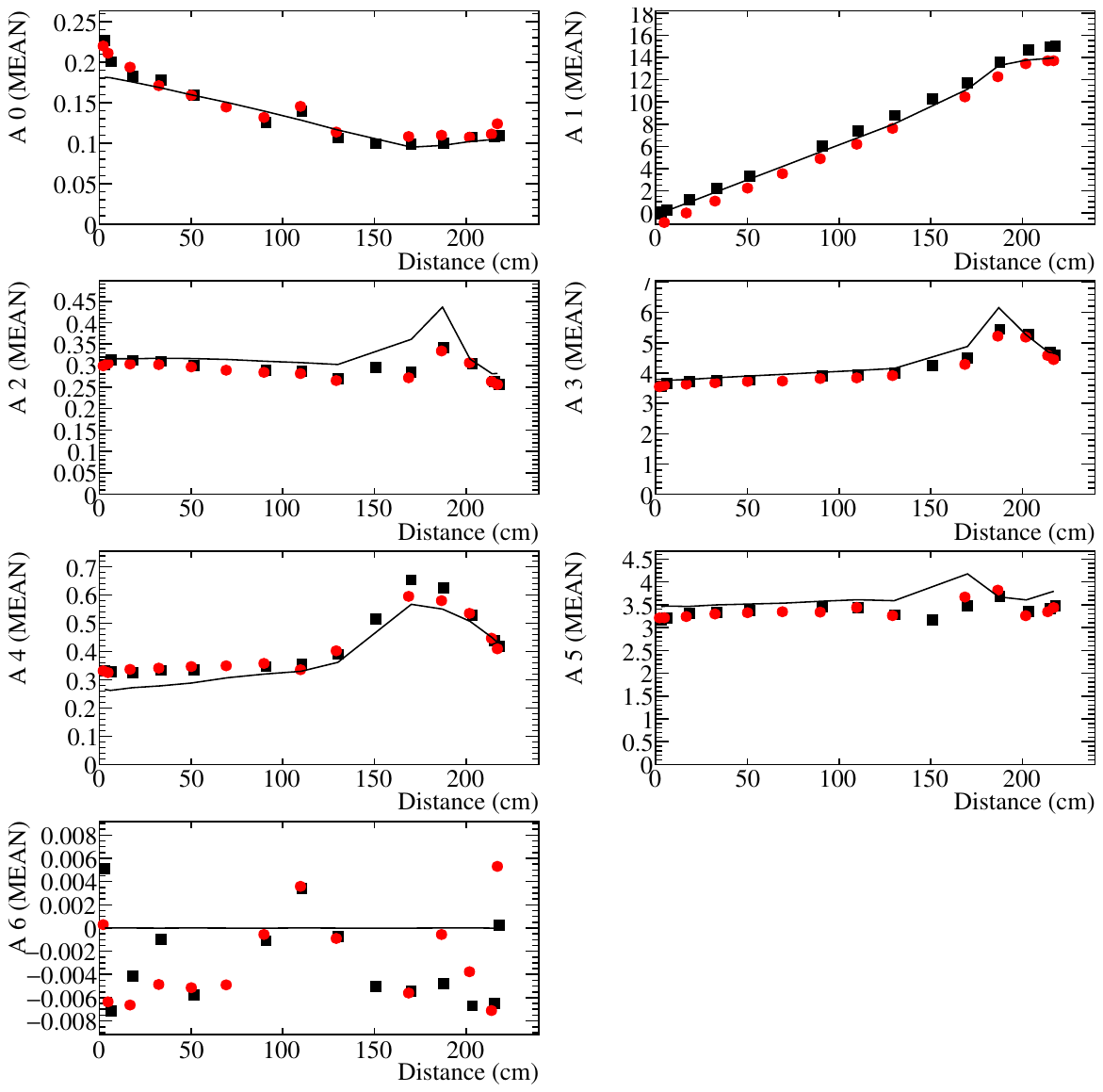}
    \caption{The mean values of the waveform fit parameters $A_0$ to $A_6$  (see Eq.~\ref{eq:WaveFormFit}) are shown, with the simulation results as a continuous line and the two detector ends in red and black, respectively. One measurement point is missing, namely black at 70~cm and red at 150~cm.}
    \label{fig:DataSimWFMean}
        \label{fig:A_mean}  
\end{figure}

\begin{figure}
    \centering
        \includegraphics[width=\linewidth]{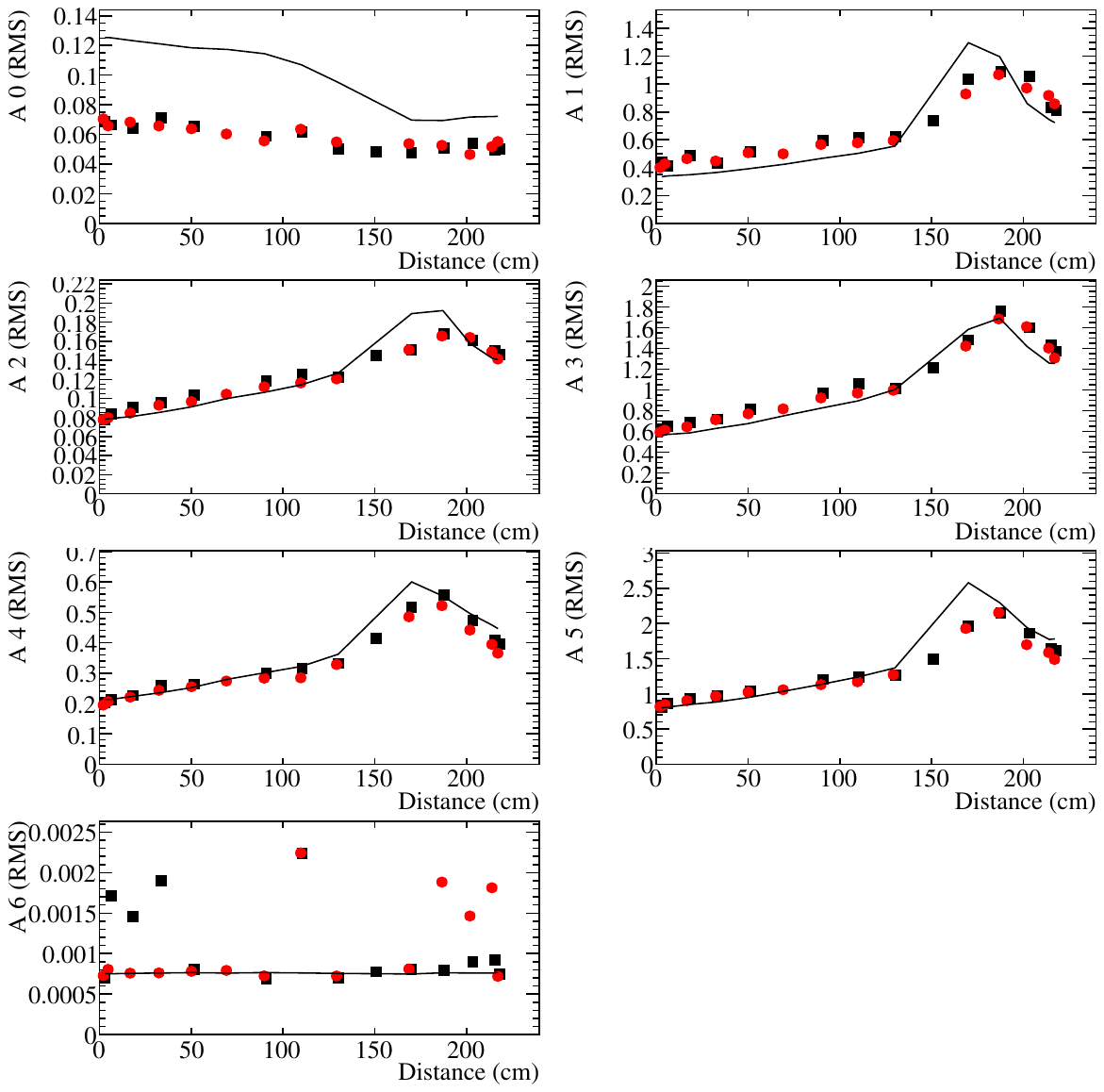}
    \caption{The RMS values of the waveform fit parameters $A_0$ to $A_6$  (see Eq.~\ref{eq:WaveFormFit}) are shown, with the simulation results as a continuous and the two detector ends in red and black, respectively. One measurement point is missing, namely black at 70~cm and red at 150~cm.}
    \label{fig:DataSimWF_RMS}
    \label{fig:A_rms} 
\end{figure}

\begin{figure}
    \centering
    \includegraphics[width=0.87\linewidth]{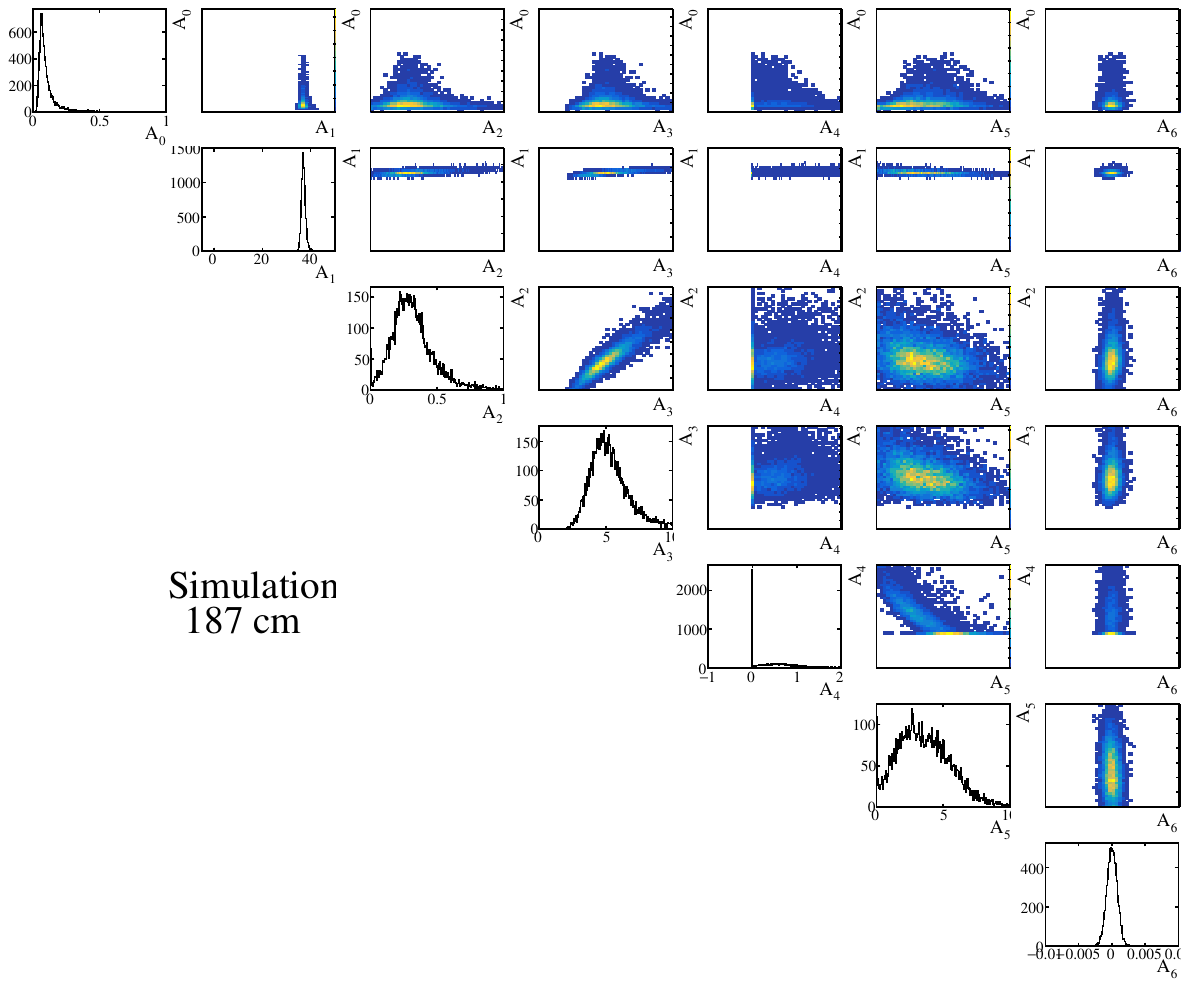}
   \includegraphics[width=0.87\linewidth]{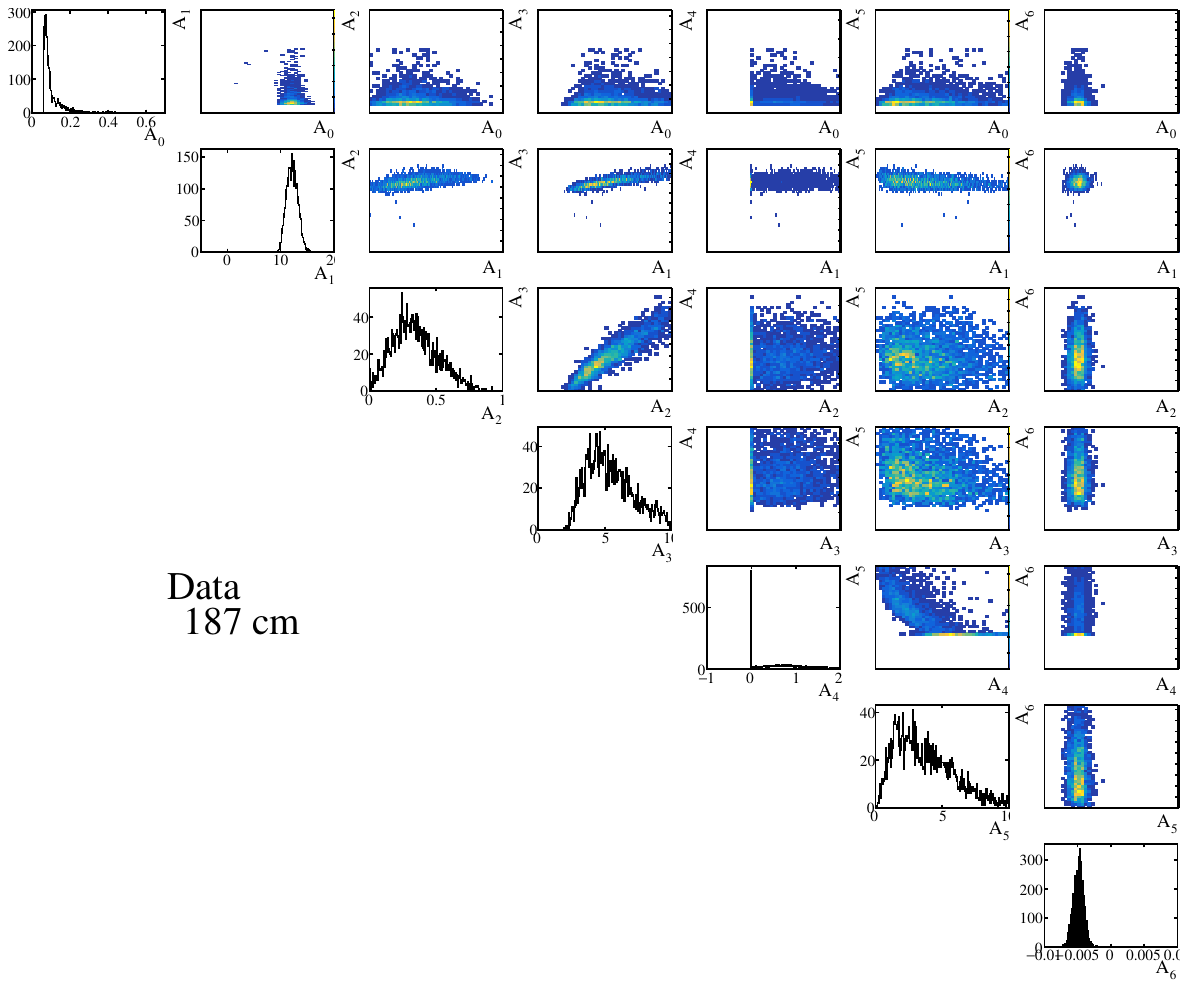}
    \caption{Waveform parameter fit results for the single bar simulation at a distance of 187~cm. Top matrix is obtained with the simulation and the lower matrix is data from the single bar setup. }
    \label{fig:Fit2D}
\end{figure}

The mean and RMS of the fitted $A_n$ values obtained at different positions along the single-bar setup are shown in Figs.~\ref{fig:A_mean} and \ref{fig:A_rms}. A comparison with the simulation predictions is discussed in Sec.~\ref{Sec:DataSimComp}. The results are consistent independent of the readout side.
The correlation between the various $A_N$ parameters is illustrated in Fig.~\ref{fig:Fit2D}. It is observed that parameters $A_2$ and $A_3$ exhibit a strong positive correlation, whereas $A_4$ and $A_5$ demonstrate a strong anti-correlation. This behaviour is an artefact of the definition of the modified Gaussian width, where $t - A_1$ is always negative for $t < A_1$ and always positive for $t > A_1$. In both cases, if one width parameter increases the other has the tendency to decrease. The rest of the parameter show a mild or null correlation. We can observe that none of the parameters show a large correlation with $A_0$ pointing to a good decoupling between the amplitude and the shape of the signal. 

\begin{figure}
    \centering
    \includegraphics[width=0.75\linewidth]{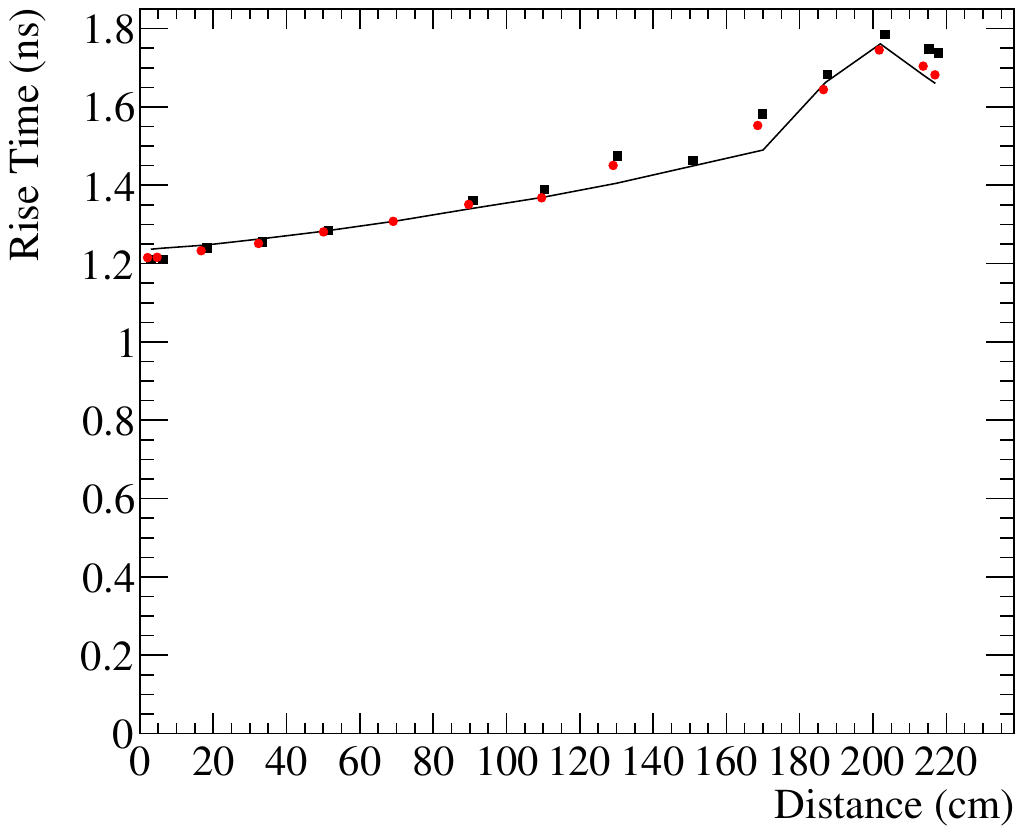}
    \caption{The rise time, computed as the difference between the time at 10\% of the signal and at 40\% of the signal, of the waveform acquired by the photo-sensors. The two side readouts are shown independently (red dots and black squares). The line shows the values predicted by the model. }
    \label{fig:RiseTime}
\end{figure}

Based on the waveform fitting results, we can calculate the signal's rise time and its dependence on the distance to the sensor. The difference between the times obtained with $\tau = 10\%$ and $\tau = 40\%$ threshold values is shown in Fig.~\ref{fig:RiseTime}.
The rise time increases slightly and linearly up to 160~cm. This increase is associated with the average time of photons reaching the sensor, which increases with distance due to angular dispersion. Beyond 160~cm, there is a change of the slope as a result of the superposition of light reflected from the opposite end to the sensor as it will be described below. Starting from 180~cm, the value begins to decrease because the arrival times of the incident and reflected light start to overlap.

\section{Particle Position and Time Reconstruction}
\label{Sec:ParticlePositionAndTimeReconstruction}
The two end readouts allow for the reconstruction of the coordinates of the entry position of the track inside the bar. This reconstruction is based on the difference in time determinations at the bar ends ($t_1$ and $t_2$). The time of arrival at one end can be computed as follows:
\begin{eqnarray}
 t_{L} &=& t_0 + \frac{d}{v}  + \delta(d) \\
 t_{R} &=& t_0 + \frac{L-d}{v} + \delta(L-d) 
\label{eq:timeArrival}
\end{eqnarray}
where, $t_0$ is the arrival time of the track to the detector, $d$ is the distance of the entry point to the end of bar and $L$ is the total length of the detector and $v$ the average propagation velocity of the scintillator light inside the plastic bar. $\delta(d)$ accounts for the time walk in the reconstruction of the time due to the change in the shape of the waveform as function of the distance, $d\ge0$ or $(L-d)\ge0$, to the photo-sensor. 

The time walk effect can be attributed to the increased time it takes for photons to reach the sensor as the distance to the sensor increases. Photons will be traversing different distances $d_t$ depending on the angle of the emission of the photon $\theta$: 
\begin{eqnarray}
  d_{t} = \frac{ d } {\cos{\theta}} 
\end{eqnarray}
where $d$ is the distance from the photon emission point to the readout sensor, and $\theta$ is the azimuthal angle of emission relative to the bar axis.

Photons will arrive then with a time spread caused by the angles, from :
$ d/v$ to $(d/v)/\cos{\theta_{c}}$, where $\theta_{c}$ is the maximum angle undergoing full reflection. The spread of the photon time arrival $\delta$ would depend on the distance to the photo-sensor.  
\begin{eqnarray*}
\delta \approx \frac{d}{v} \left(1 - \frac{1}{\cos{\theta_{c}}} \right) 
\end{eqnarray*}
Since $\theta_{c}$ is constant, the dependency should be linearly proportional to the distance $d$. However, due to the limited photon statistics, the spread of photon arrival times will result in an increased dispersion that departs from simple linear dependency. 

The position and time crossing of a particle with the detector can be obtained as : 
\begin{eqnarray}
 d &=& \frac{v}{2} \left(t_L-t_R + \frac{L}{v}\right)+ \frac{v}{2}\left( \delta(L-d) -\delta(d) \right)\\
 t_0 &=& \frac{1}{2} \left(t_L+t_R-\frac{L}{v}\right) -  \frac{1}{2}\left(\delta(d)+\delta(L-d)\right) 
\label{eq:positionandtime}
\end{eqnarray}
It is important to note that the time walk functions appear with the same sign for the determination of absolute time and they are subtracted for the position determination. The time walk value can be obtained from the rise time of the signals (see Fig.~\ref{fig:RiseTime}). This rise time is almost linear with the distance to the sensor and it has a different sign for each end. 
The observed increase in rise time near 220 cm originates from the enhanced contribution of reflected photons: at first, their delayed arrival broadens the signal and lengthens the rise time, whereas at larger distances, when the reflected and direct photons nearly overlap in time, their superposition leads to a reduction of this value.

\begin{figure}
    \centering
    \includegraphics[width=0.75\linewidth]{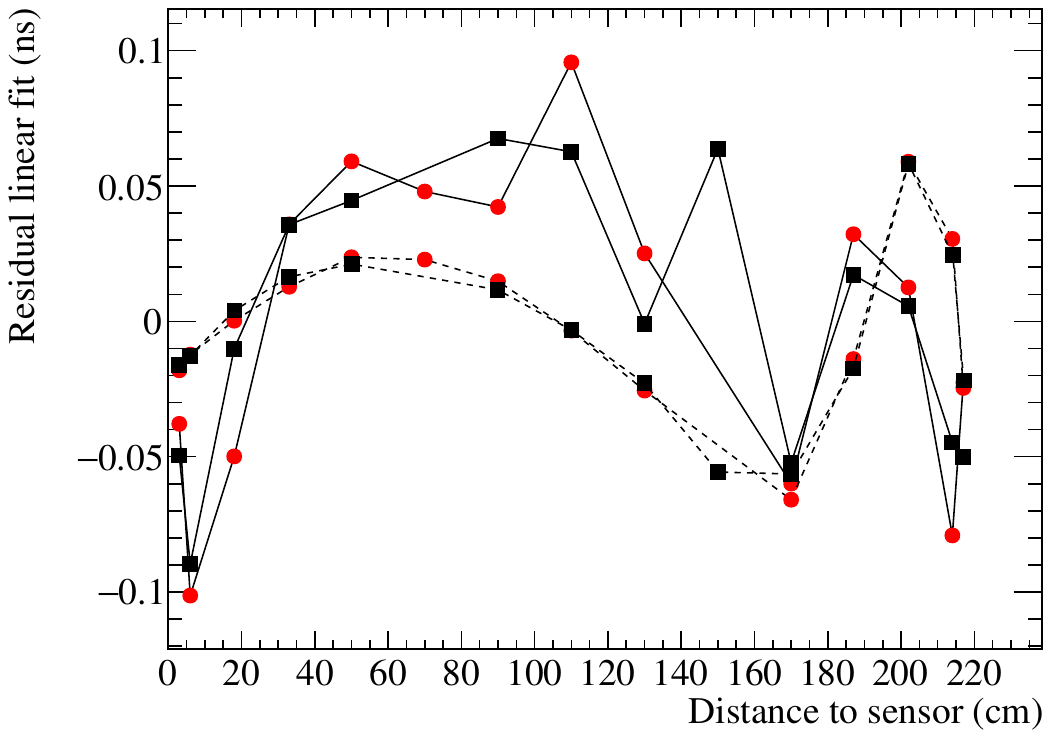}
    \caption{Residuals of the average photon arrival time as a function of distance, shown for the left (black) and right (red) sides. Solid lines correspond to measurements with a single bar, while dashed lines denote the simulation. The residuals are evaluated after subtraction of a linear fit.}
    \label{fig:SiPM-PMT}
\end{figure}

To estimate the value of $\delta(d)$, we rely on the single bar studies. The linear-fit residuals of the difference between the bar time at a given position and that of the reference photo-multiplier can be represented as a function of distance. Since the signal propagation time in the bar depends on the distance to the sensor, we correct for this effect by assuming a constant velocity:
\begin{equation}
\delta(d) \approx \Delta t_{L,R} = t_{L,R} - t_{\text{ref}} - \frac{d}{v_{\text{eff}}},
\end{equation}
where $v_{\text{eff}}\approx16$~cm/ns is an effective velocity extracted from the linear dependence on distance, and $t_{\text{ref}}$ is the reference time provided by an external source (e.g., a photomultiplier in the setup or an offset in the simulation). The photon propagation velocity is an effective parameter incorporating both transport time and additional effects such as reflection and scattering, which cause a dispersion of photon arrival times at the sensor. Effective velocities are obtained from fits to both data and simulation and are discussed in Sec.~\ref{Sec:DataSimComp}. The fit residuals,
\begin{equation}
\Delta t_{L,R} + \frac{d}{v_{\text{eff}}}
\end{equation}
is shown in Fig.~\ref{fig:SiPM-PMT}. It exhibits a complex structure, with clear deviations at the bar ends, near the readout sensor, and around 170 cm, where reflected photons begin to overlap with direct ones. Consequently, we expect that the term
\begin{equation}
\delta(d) \pm \delta(L - d)
\end{equation}
remains visible at the order of 100~ps in both the $t_0$ and position determination.  

\subsection{Time determination with external track coordinate}
\label{Sec:TimeDetermExt}
Since we use the measurements on each side of the bar to obtain coordinates and momentum, the time resolution is not optimal. The optimal resolution would be achieved by determining, with other detectors (such as SuperFGD~\cite{Blondel:2020hml} and HA-TPC~\cite{Attie:2022smn} in ND280), the entry point of the track ($d_0$), and then utilising the two independent time measurements to optimise time determination.
\begin{eqnarray}
 t^{(L)}_{0} &=& t_L - \frac{d_0}{v}  - \delta(d_0) \\
 t^{(R)}_{0} &=& t_R - \frac{L-d_0}{v}  - \delta(L-d_0)
\label{eq:timeArrival2}
\end{eqnarray}
 The time with the optimal resolution $t^{m}_{0}$ is then obtained as the weighted averaged :
\begin{eqnarray}
 t^{m}_{0} &=& \frac{\frac{ t^{(L)}_{0} }{\sigma^2_{t_L}}+\frac{t^{(R)}_{0} }{\sigma^2_{t_R}}} {\frac{1}{\sigma^2_{t_L}}+\frac{1}{\sigma^2_{t_R}}}
\end{eqnarray}
with $\sigma_{t_{L,R}}$ being the error associated to the time determination at both bar ends (Left and Right). These errors depend on the distance from the track to the sensor \cite{Korzenev:2021mny}. It is important to notice that this measurements requires accurate determination of the entry position as well as the knowledge of the time walk dependencies $\delta(d)$ and the nominal propagation velocity $v$. 

\section{Comparison with data}
\label{Sec:DataSimComp}
\subsection{Muon flux model and energy deposition}
To compare with the single bar results, we made a simulation that takes into account the cosmic ray muon energy distribution for vertically incident muons, the angle is forced by the external trigger used in the single bar setup. The muon energy spectrum is given by the Gaisser reparametrisation for $\theta = 0$~\cite{PDG2022} :
\begin{equation}
\frac{d \phi_{\mu}}{d E_{\mu}} \propto E_{\mu}^{-2.7} \left(
\frac{1}{1 + 1.1 \, \frac{E_{\mu}}{115\,\text{GeV}}}
+ 
\frac{0.054}{1 + 1.1 \, \frac{E_{\mu}}{850\,\text{GeV}}}
\right) \text{GeV}^{-1}
\end{equation}

For a given muon energy, we model the energy deposition in the scintillator based on a simplified Bethe-Block formula~\cite{PDG2022},
\begin{equation}
\left( -\frac{dE}{dx} \right) \propto \frac{1}{\beta^2} 
\left[
\frac{1}{2} \ln\left( \frac{2 m_e \beta^2 \gamma^2 W_\text{max}}{I^2} \right)
- \beta^2 - \frac{1}{2} \delta(\beta\gamma) 
\right] \left(\frac{\text{MeV}}{\text{cm}}\right)
\end{equation}
where:
\begin{equation}
  W_\text{max} = \frac{2 m_e \beta^2 \gamma^2}{1 + 2\gamma \frac{m_e}{m_\mu} + \left( \frac{m_e}{m_\mu} \right)^2} 
\end{equation}
with $I$ the mean ionization potential, $m_e$ ($m_{\mu}$) the electron (muon) mass. The density correction $\delta$ is approximately given as:
\begin{eqnarray}
\delta(\beta\gamma) =
\begin{cases}
0, & \log{\beta\gamma} < X_0 \\[6pt]
2 (\log{\beta\gamma} - C) + D (X_1 - \log{\beta\gamma})^3, & X_0 \leq \log{\beta\gamma} < X_1 \\[6pt]
2 (\log{\beta\gamma} - C), & X_1 \leq \log{\beta\gamma}
\end{cases}
\end{eqnarray}
where for plastic scintillators, the parameters are
$X_0 = 0.46$, $X_1 = 4.6$, $C = 6.9$ and $D = 0.0082$.
The Bethe-Bloch ionisation value is normalised to obtain the typical $2$~MeV/cm~\cite{PDG2022} at the minimum of the ionisation. 

We describe the fluctuation of the dE/dx in a 1~cm plastic (straggling) with a Landau distribution with mean energy deposition value provided by the Bethe-Bloch formula and a sigma equal to 20\% of this value \cite{Jones1968}.

\subsection{Data Model comparison }

\begin{figure}
\centering
\includegraphics[width=0.75\linewidth]{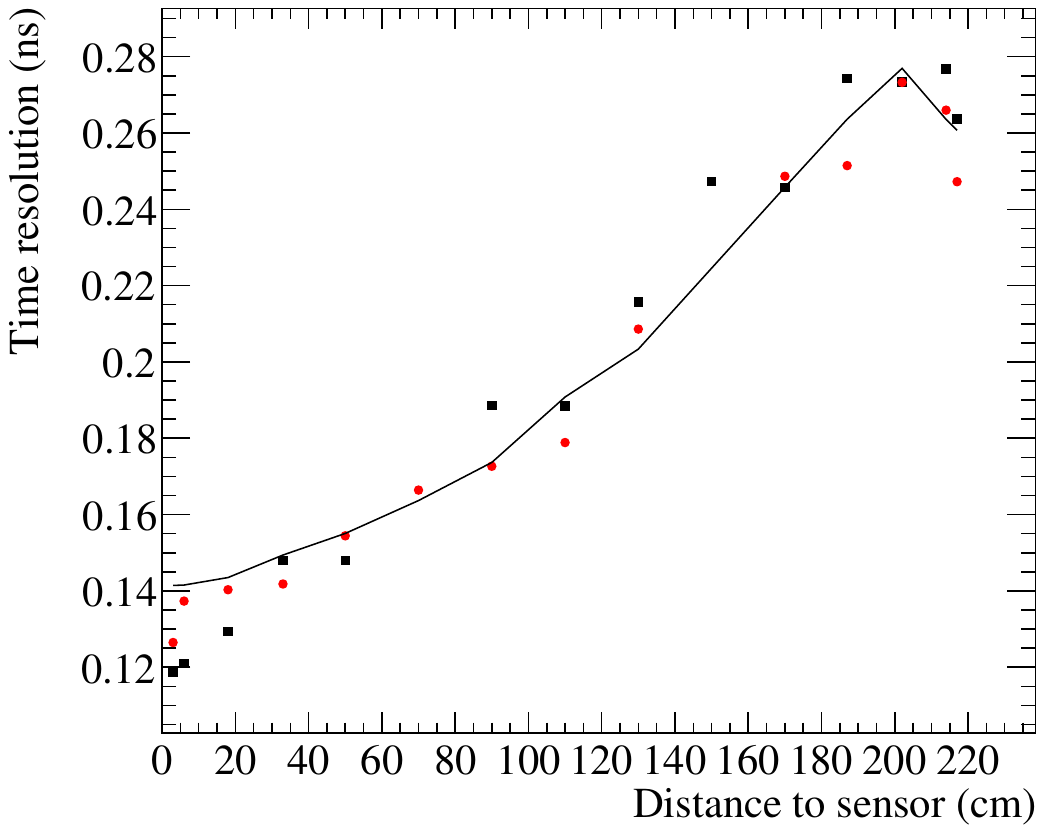}
\caption{Time resolution obtained from the detector model (continuous line) compared to the single-bar data (red and black stands for the left and right detector ends). }
\label{fig:DataSim}
\end{figure}

We have used the parameters listed in Table~\ref{tab:SimParam} to compare simulation results with experimental data, as illustrated in Fig.~\ref{fig:DataSim}. The agreement in resolution is excellent. Fig.~\ref{fig:DataSimWFMean} and~\ref{fig:DataSimWF_RMS} present a comparison of the waveform parameters extracted from data and simulation. While the agreement here is less precise, the discrepancies do not significantly impact the overall results.

Focusing on the mean values of the waveform parameters, defined in Eq.~\ref{eq:WaveFormFit} as the $A_x$ parameters, the largest differences occur for $A_3$, $A_4$, $A_5$, and $A_6$, with a smaller deviation observed for $A_2$. Specifically, $A_6$ represents the baseline shift, and $A_4$ and $A_5$ characterize the falling edge of the waveform. Among these parameters, only $A_2$ and $A_3$ are expected to influence the time resolution. Their values are systematically higher in the simulation than in the data; however, this difference does not appear to significantly affect the time resolution, as shown in Fig.~\ref{fig:DataSim}.

\begin{figure}
\centering
\includegraphics[width=.5\linewidth]{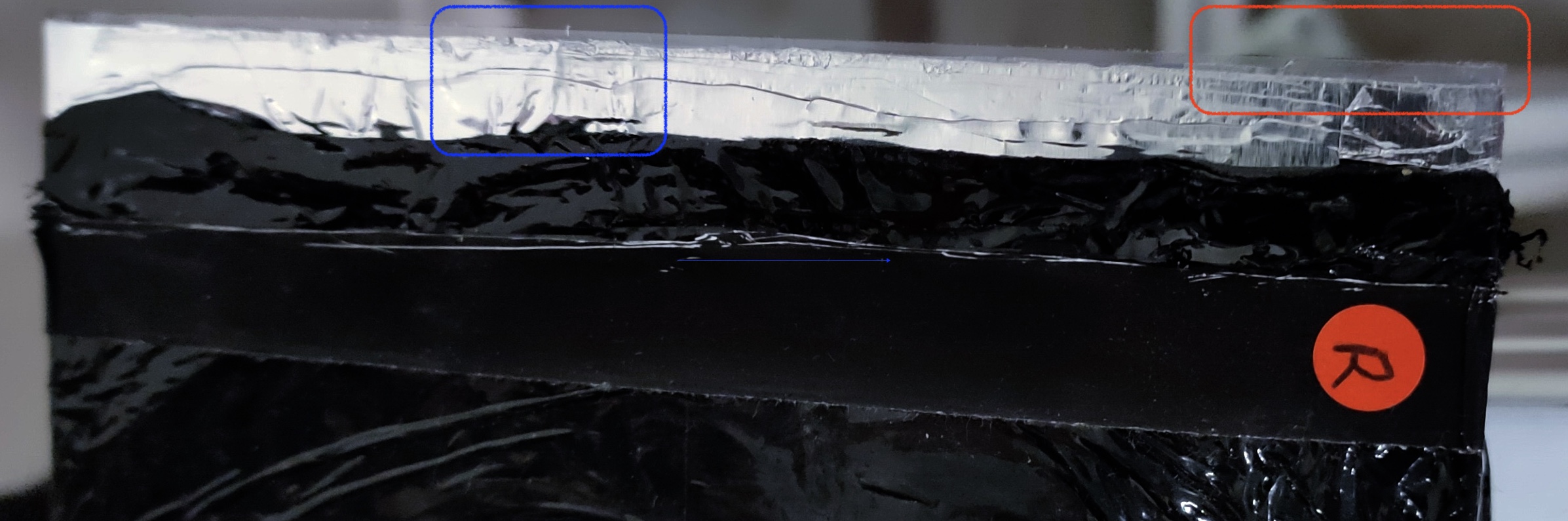}
\includegraphics[width=.5\linewidth]{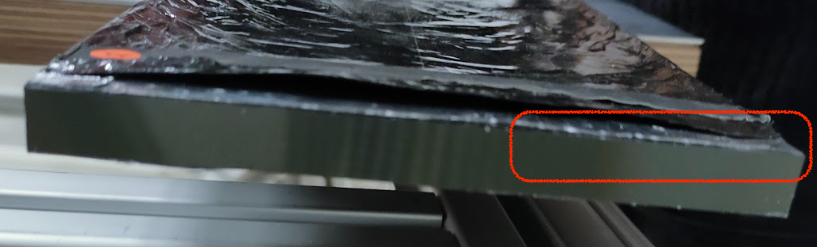}
\caption{End of the scintillator bar near the readout capsule. Regions with deficient wrapping are highlighted by red boxes, while areas with a corrugated wrapping are indicated by blue boxes.}
\label{fig:EndofBar}
\end{figure}

The mean value of $A_6$ in the data exhibits significant variation, which is attributed to changes in the ambient temperature during the data acquisition period, which lasted several months. The SiPM is very sensitive to temperature variations, which can significantly affect the baseline noise level characterized by $A_6$. We do not observe a similar dependency for the other parameters $A_i$, indicating that the reconstruction correctly accounts for baseline level shifts.
The model does not perfectly reproduce the bump observed at distances greater than 160~cm for parameters $A_2$ to $A_4$. These bumps originate from reflected photons. The discrepancy primarily affects distances beyond 180~cm, indicating a deficit of reflected photons in this region. A similar effect is also observed in the parameter $A_0$. Adjusting this region with the current model is challenging, particularly because the agreement above 1.8~m is reasonable. This discrepancy is most likely associated with edge effects of the bar that are not included in the simulation, such as the connection of the readout module to the bar. These imperfections can reflect photons before they reach the end of the bar or alter the photon direction, thereby affecting reflections at the readout end. As shown in Fig.~\ref{fig:EndofBar}, the aluminium cover at the end of the bar, where the readout capsule is connected, is not perfectly flat and is sometimes partially missing.

The Root Mean Square (RMS) of the parameters $A_i$ is shown in Fig.~\ref{fig:A_rms}. While the agreement is slightly worse than for the mean values, it remains reasonable given the simplicity of the model. A notable discrepancy appears in the RMS of $A_0$, but this is not critical, as it primarily reflects the simulation of the energy deposition. This likely results from the model's simplified treatment of the muon energy spectrum and energy deposition, which may be influenced by the adopted 20\% straggling value. Another parameter not well reproduced is the RMS of $A_6$, which, like its mean, is affected by temperature variations along the data acquisition period. The shape of the RMS of $A_6$ closely resembles that of the $A_6$ mean. The shapes of the RMS for the remaining parameters are generally well reproduced, with similar difficulties in describing the transition from direct to direct-plus-reflected light pulses.

The simulation also captures the correlations between the different $A_i$ parameters, as illustrated in Fig.~\ref{fig:Fit2D}\footnote{In the data, $A_1$ is shifted due to the lack of proper zero suppression.}. Among these, $A_5$, which describes the waveform’s falling edge, shows the largest deviation in the simulation. However, since $A_5$ has minimal impact on resolution or rise-time calculations, this discrepancy has limited practical consequences.

In order to achieve this level of agreement, we have introduced a slight increase in the effective reflectivity of the 3D-printed material surrounding the SiPM encapsulation (see Table~\ref{tab:SimParam}). This adjustment can be interpreted as compensating for an incomplete modelling of the complex optical response at the end of the detector as it was discussed above. 
Furthermore, the rise time comparison shown in Fig.~\ref{fig:RiseTime} demonstrates excellent agreement across all propagation distances. A small discrepancy remains in the rise time between data and simulation for distances between 160~cm and 180~cm; this is the region where direct and reflected light pulses at the end of the bar start to overlap in time. The same behaviour is observed in the mean values of parameters $A_3$ and $A_4$, which are also sensitive to the relative contribution of reflected light.

\begin{figure}
\centering
\includegraphics[width=0.75\linewidth]{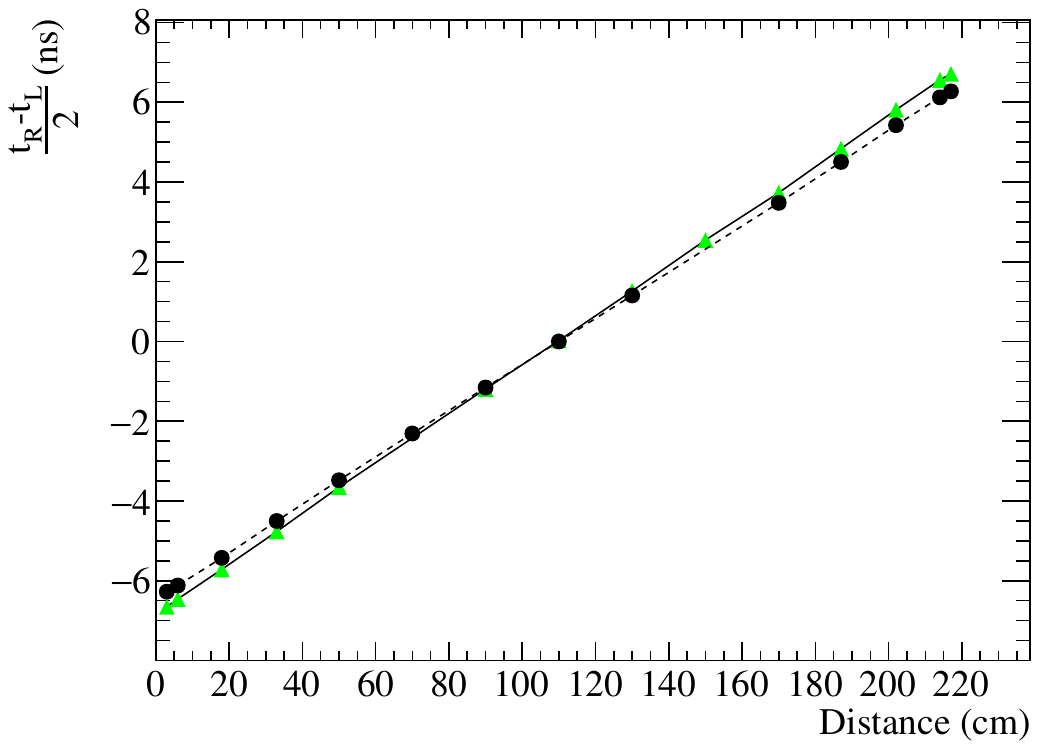}
\caption{ Estimation of the velocity for data (green triangles) and simulation (black squares with dashed lines). The velocity is estimated from the difference of the arrival time of the signal to both detector ends. }
\label{fig:Speed}
\end{figure}

The photon propagation velocity is an effective parameter that accounts not only for the photon propagation itself but also for attenuation and interactions. To evaluate the effective velocity, we compute half the difference in the average arrival time at both ends, as described in Eq.~\ref{eq:timeArrival}, assuming the value of $\delta(d)$ to be negligible:
\begin{equation}
\frac{t_R - t_L}{2} \approx \frac{d}{v} - \frac{L}{2v}.
\end{equation}
We present this quantity for both data and simulation in Fig.~\ref{fig:Speed}. The inverse of the slope serves as an estimator of the velocity. Our measurements yield a propagation velocity of 16~cm/ns in the single-bar data, which is lower than both the value of 19~cm/ns inferred from the tabulated refractive index of EJ-200 and the result obtained with this index of refraction and the photon propagation simulation 17~cm/ns. While not conclusive, this discrepancy can be explained by noting that the tabulated value typically corresponds to the phase refractive index ($n_p$), whereas the transport of an optical pulse is governed by the group refractive index ($n_g$). The relationship between these indices is given by:
\begin{equation}    
n_{g}(\lambda) = n_{p}(\lambda) - \lambda \frac{dn_{p}(\lambda)}{d\lambda}.
\end{equation}
In the visible range, the derivative $\frac{dn_p}{d\lambda}$ is typically negative due to normal dispersion (the refractive index $n_p$ decreases with increasing wavelength $\lambda$). Consequently, the group index exceeds the phase index, causing an optical pulse to propagate more slowly than the monochromatic phase velocity at a given wavelength.

Achieving a more accurate simultaneous reproduction of both the resolution and the waveform shape will require further optimization of the simulation parameters, as well as a more realistic modelling of the particle momentum and angular distributions. The latter can only be achieved within the framework of the full ND280 T2K Monte Carlo. Nevertheless, the present level of agreement already represents a significant milestone for T2K operations: it enables the use of simulation outputs in a manner directly comparable to experimental data, while retaining a detector response that is modelled from first principles. This capability is particularly important for future evaluations of systematic uncertainties in timing measurements based on time-of-flight (ToF).

\section{Conclusions}

We have introduced an algorithm to parametrise waveforms obtained with the ND280 upgrade Time-of-Flight detector. This technique enables optimal time determination using a constant-fraction threshold method and allows the study and monitoring of waveform shapes with only a few parameters. We also present a model for photon transport and detection that accurately reproduces the detector response using basic input parameters. The simulation’s resolution is consistent with that of the single-bar calibration setup, and the waveform shapes, parametrised by the 
$A_n$ parameters, show excellent agreement. The model is computationally efficient, enabling straightforward evaluation of potential geometrical or material modifications, and can be readily adapted to study the performance of similar detectors, thereby simplifying the design phase.


\appendix

\acknowledgments

The research was supported and funded by the Swiss-Vietnam project: Science Nation Foundation, Switzerland and by Vietnam National Foundation for Science and Technology Development (NAFOSTED), under grant number  IZVSZ2.203433. We gratefully acknowledge the support of the Science and Technology Facilities Council (STFC) and UK Research and Innovation (UKRI), United Kingdom.


\bibliographystyle{JHEP}
\bibliography{biblio.bib}

\end{document}